\documentclass[a4paper,11pt]{article}
\usepackage{pos}
\usepackage{wrapfig}
\graphicspath{{Figures/}}

\newcommand{\gthdm}{\textit{g}2HDM}
\newcommand{\order}[1]{\mathcal{O}(#1)}
\renewcommand{\Re}{\mathrm{Re}}
\renewcommand{\Im}{\mathrm{Im}}

\renewcommand{\acknowledgments}{\noindent\textbf{Acknowledgements. }}

\title{Lepton and neutron EDM as probe of general two Higgs doublet model}
\ShortTitle{eEDM and nEDM in g2HDM}

\author{Sven Teunissen}

\affiliation{National Taiwan University,\\
  1 Roosevelt Road Sec. 4, Da'an District, Taipei 10617, Taiwan, R.O.C.}

\emailAdd{r09222070@ntu.edu.tw}

\abstract{Baryogenesis requires large CPV phases, while said phases are constrained by electric dipole moment (EDM) experiments.
In the \textit{general} two Higgs doublet model ({\gthdm}), without \(Z_{2} \) symmetry, EWBG can be achieved while evading EDM bounds.
In this study, we explore the {\gthdm} contributions to electron EDM (eEDM) and neutron EDM (nEDM), and review the future prospects in the experiment front.
In particular, we show that the combined eEDM-nEDM results can not only provide crucial bound on the top Yukawa-driven baryogenesis explanation in {\gthdm}, but are \textit{poised for discovery} as experimental precision increases within the next decade or so.}

\FullConference{The European Physical Society Conference on High Energy Physics (EPS-HEP 2023)\\
 21-25 August 2023\\
Hamburg, Germany\\}

\renewcommand{\logo}{\relax}

\begin{document}
\maketitle

\section{Introduction}
One of the biggest unanswered questions of particle physics is that of baryogenesis.
Specifically, if electroweak baryogenesis (EWBG)~\cite{EWBG} were to occur, one would require very large CP violation (CPV) beyond the Standard Model (BSM), since the SM currently houses all its CPV in the CKM matrix~\cite{PDG}.
However, such large BSM-CPV should have led to new discoveries at the LHC, which evidently is \textit{not} what has been observed.
Moreover, in the low-energy precision frontier, electric diploe moments (EDMs) provide a \textit{litmus test} for CPV effects, and experiments have achieved higher and higher precision without discoveries, setting ever more stringent bounds.
In a sense, these ``tabletop experiments'' are directly competing with the LHC!

\section{The General 2HDM}
We do not state and discuss the Higgs potential for {\gthdm} here, but instead focus directly on the Lagrangian and its flavor characterisics. The {\gthdm} Lagrangian can be written as~\cite{DavidsonHaber05, HouModak21}
\begin{align}
  \mathcal{L} = - & \frac{1}{\sqrt{2}} \sum_{f = u, d, \ell} \bar f_{i} \Big[\big(-\lambda^f_i \delta_{ij} s_\gamma + \rho^f_{ij} c_\gamma\big) h
  + \big(\lambda^f_i \delta_{ij} c_\gamma + \rho^f_{ij} s_\gamma\big)H
  - i\,{\rm sgn}(Q_f) \rho^f_{ij} A\Big]  R\, f_{j} \nonumber                                                                                     \\
  -               & \bar{u}_i\left[(V\rho^d)_{ij} R-(\rho^{u\dagger}V)_{ij} L\right]d_j H^+
  - \bar{\nu}_i\rho^L_{ij} R \, \ell_j H^+ + {\rm h.c.},
  \label{eq:lagrangian}
\end{align}
where the generation indices \(i \), \(j \) are summed over, \(L \), \(R = (1\pm\gamma_{5})/2\) are projections, \(V \) is the CKM matrix for quarks and unity for leptons.
\(\lambda^f \) are the SM Yukawa matrices, and \(\rho^f \) are the extra-Yukawa matrices.
A key takeaway is that each family of fermions (u-type, d-type, lepton) is associated with its own extra-Yukawa \(\rho \) matrix.
In this scenario, flavor-changing neutral Higgs (FCNH) processes are controlled by \textit{flavor hierarchies} and \textit{alignment}.
Flavor hierarchies means that the \(\rho \) matrices somehow ``know'' the current flavor structure of the SM, represented by the ``rule of thumb''~\cite{HouKumar20}
\begin{equation}
  \rho_{ii} \lesssim \order{\lambda_i}, \quad
  \rho_{1i} \lesssim \order{\lambda_1}, \quad
  \rho_{3j} \lesssim \order{\lambda_3},
  \label{eq:ruleofthumb}
\end{equation}
with \(j\neq 1 \).
Alignment means that \(c_{\gamma} \equiv \cos\gamma = \cos(\beta-\alpha)\) is small.
Consequently, the SM-like Higgs \(h \) is mostly controlled by the SM Yukawas, while the newly introduced \(\rho \) matrices control the exotic Higgses \(H, A, H^{\pm} \).
A remarkable feature of {\gthdm} is that \(\order{1}\,\rho_{tt}\) can drive EWBG through~\cite{FSH18} \(\lambda_{t}\Im\rho_{tt} \). This feature, however, is immediately put the test in the realm of EDMs.

\section{EDM as precision probes}
\begin{wrapfigure}{r}{0.31\textwidth}
  \centering
  \includegraphics[width=0.31\textwidth]{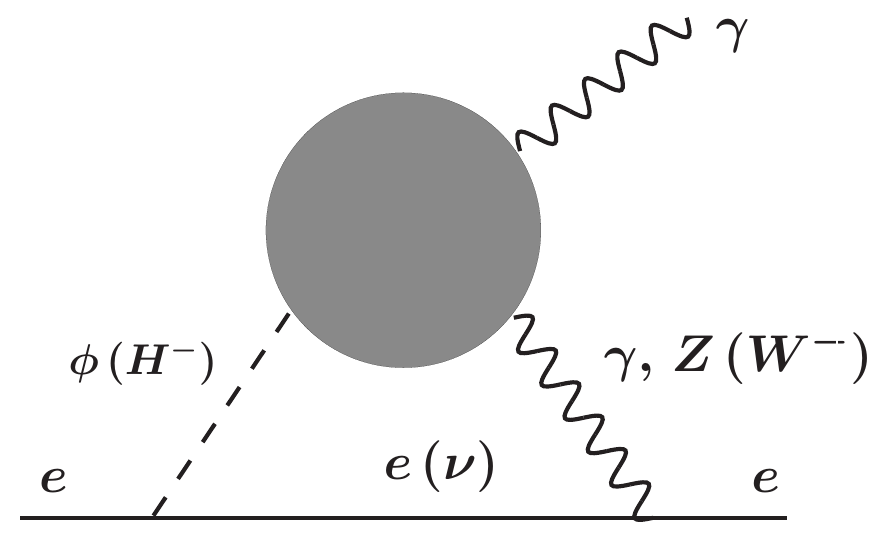}
  \caption{Two-loop Barr-Zee diagram for the electron.}
  \label{fig:barrzee}
\end{wrapfigure}
The effective interaction term that produces EDM \(d_{f} \) for a fermion \(f \) is the dimension-5 operator
\begin{equation}
  -\frac{i}{2}d_{f}\left(\bar{f}\sigma^{\mu\nu}\gamma_{5}f\right)F_{\mu\nu}.
\end{equation}
In {\gthdm} the main contribution to this operator is the two-loop Barr-Zee diagram~\cite{BarrZee90} (Fig.~\ref{fig:barrzee}).
We evaluate \(d_{f} \) following Ref.~\cite{Abe14}.
Due to the high top mass, the loop contributions are dominated by the top loop, thus \(\rho_{tt} \) becomes the main contributing parameter of {\gthdm} in EDMs (alongside the \(\rho_{ff} \) for the fermion in question, of course).
Thus, large \(\rho_{tt} \) should lead to large EDMs.

\subsection{Electron EDM}
\begin{figure}[t]
  \centering
  \includegraphics[width=5.3cm,height=3.7cm]{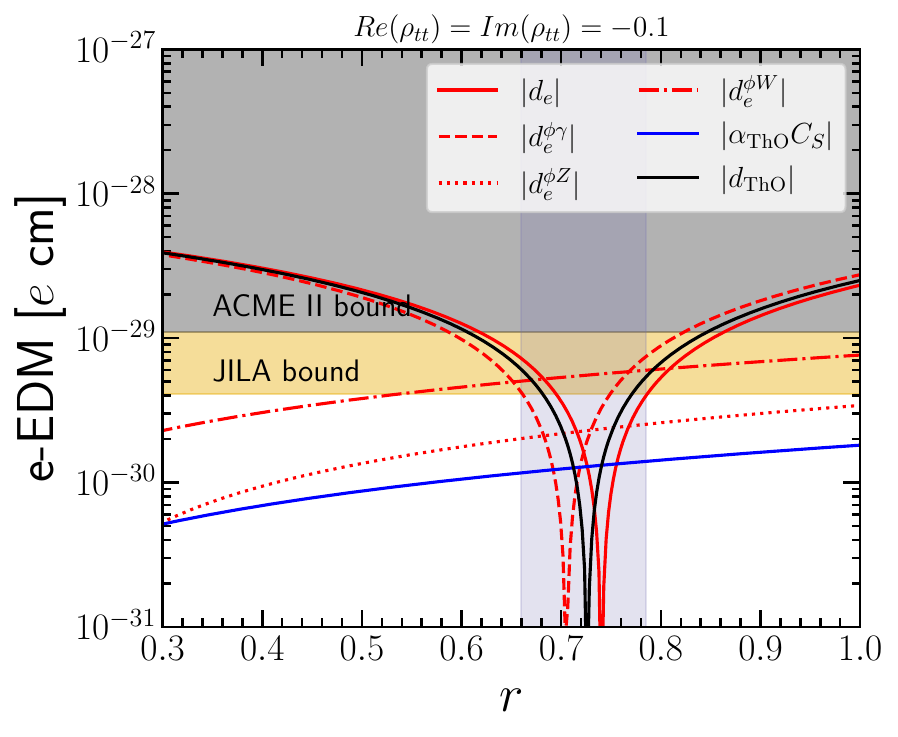}
  \includegraphics[width=4.63cm,height=3.7cm]{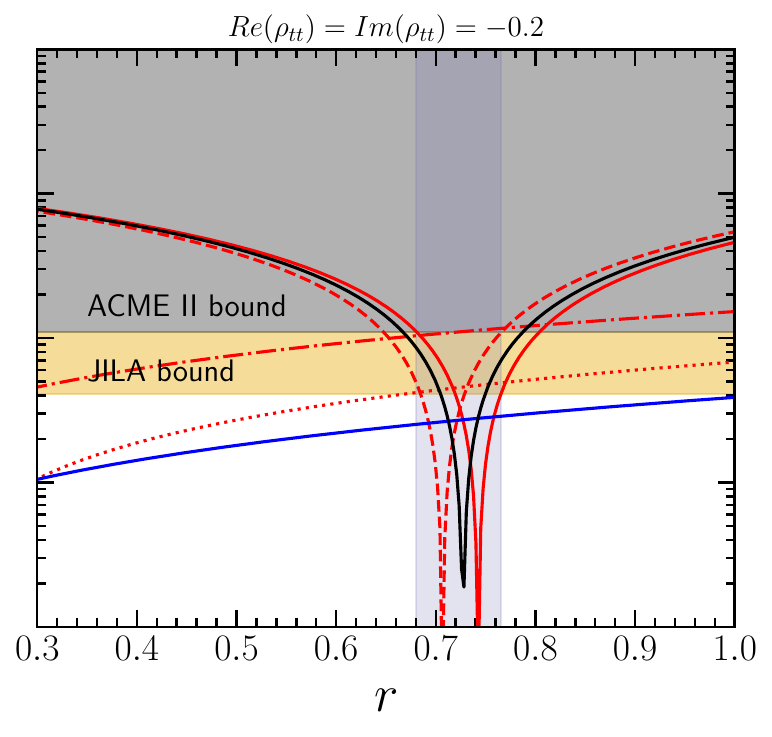}
  \includegraphics[width=4.63cm,height=3.7cm]{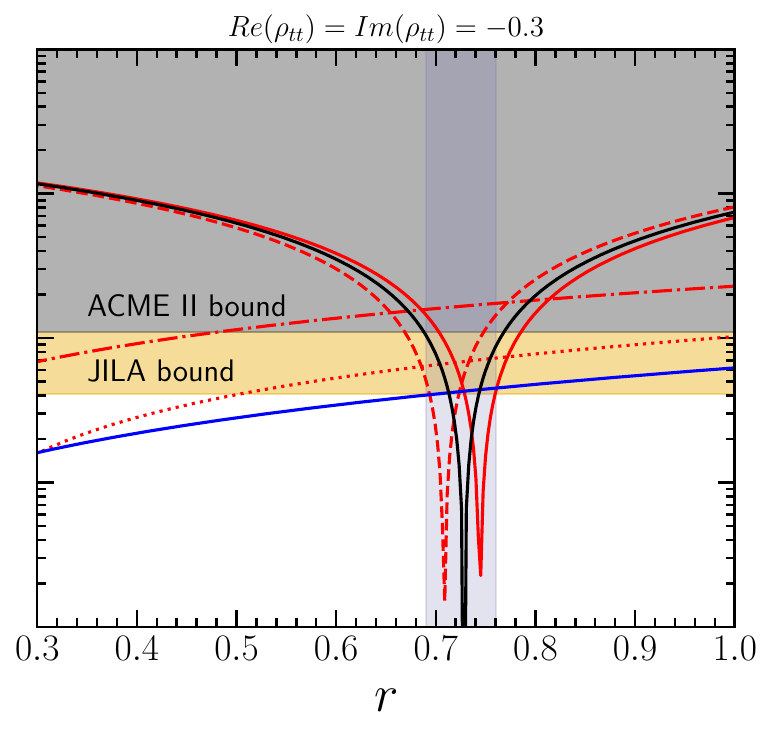}
  \caption{eEDM v.s. \(r \) for a larger range of \(\rho_{tt} \) with ansatz~Eq.~\eqref{eq:ansatz}. (\(c_{\gamma} = 0.1, m_{H, A, H^+} = 500\,\mathrm{GeV} \))}
  \label{fig:eEDM}
\end{figure}

The experimental development of electron EDM (eEDM) over the past few years has been remarkably rapid.
Just earlier this year, JILA~\cite{JILA23} has surpassed the previous bound from ACME~\cite{ACME18} and pushed the precision of eEDM down to \(|d_{e}| < 4.1 \times 10^{-30}\) \(e\,\mathrm{cm} \).
It is noteworthy to point out that these eEDM experiments are relatively small in scale, ``tabletop experiments'' even when compared to behemoths like the LHC, which makes the extreme precision achieved all the more impressive.
As mentioned above, in {\gthdm}, baryogenesis is achieved through large \(\rho_{tt} \); however, large \(\rho_{tt} \) should also yield large EDMs, yet that is not what experiments have shown us!
To address this discrepancy and evade the eEDM bounds, a previous study by Fuyuto, Senaha, and Hou~\cite{FSH20} proposed a ``cancellation ansatz'' between \(\rho_{ee} \) and \(\rho_{tt} \)
\begin{equation}\label{eq:ansatz}
  \Re\rho_{ee} = -r\frac{\lambda_{e}}{\lambda_{t}}\Re\rho_{tt} \text{, } \quad \Im\rho_{ee} = +r\frac{\lambda_{e}}{\lambda_{t}}\Im\rho_{tt},
\end{equation}
where \(r \) depends on loop functions.
Eq.~\eqref{eq:ansatz} gives both a flavor hierarchy \(|\rho_{ee}|/|\rho_{tt}|\sim\lambda_{e}/\lambda_{t} \) that reflects SM, as well as a phase lock.
In their study, they set \(\Re\rho_{tt} = \Im\rho_{tt} = -0.1 \) (which equates to \(|\rho_{tt}| = 0.1\sqrt{2} \approx 0.14\)).
In our study~\cite{HKT23-2}, we aim to explore a larger range of \(\rho_{tt} \), up to \(\Re\rho_{tt} = \Im\rho_{tt} = -0.3 \) (\(|\rho_{tt}| = 0.3\sqrt{2} \approx 0.42\)). Also, for the sake of numerical illustration of the flavor hierarchy, we extend the ansatz to all fermion \(\rho_{ff} \)s, except for the top itself. Results are shown in Fig.~\ref{fig:eEDM}.
We have taken a bit of liberty in illustrating the range of the purple ``allowed window'' band for sake of clarity.
We can see that as \(|\rho_{tt}| \) increases, the allowed window of the proportionality parameter \(r \) shrinks, yet there is still a decent range of acceptable probable values.
As a sidenote, we have also studied the EDM of muon and tau in {\gthdm} in a recent paper~\cite{HKT22}; however, the results were not as insightful, so we have chosen to leave them out of this talk.

\subsection{Neutron EDM}
\begin{figure}[t]
  \centering
  \begin{minipage}{0.48\textwidth}
    \centering
    \includegraphics[width=0.85\linewidth]{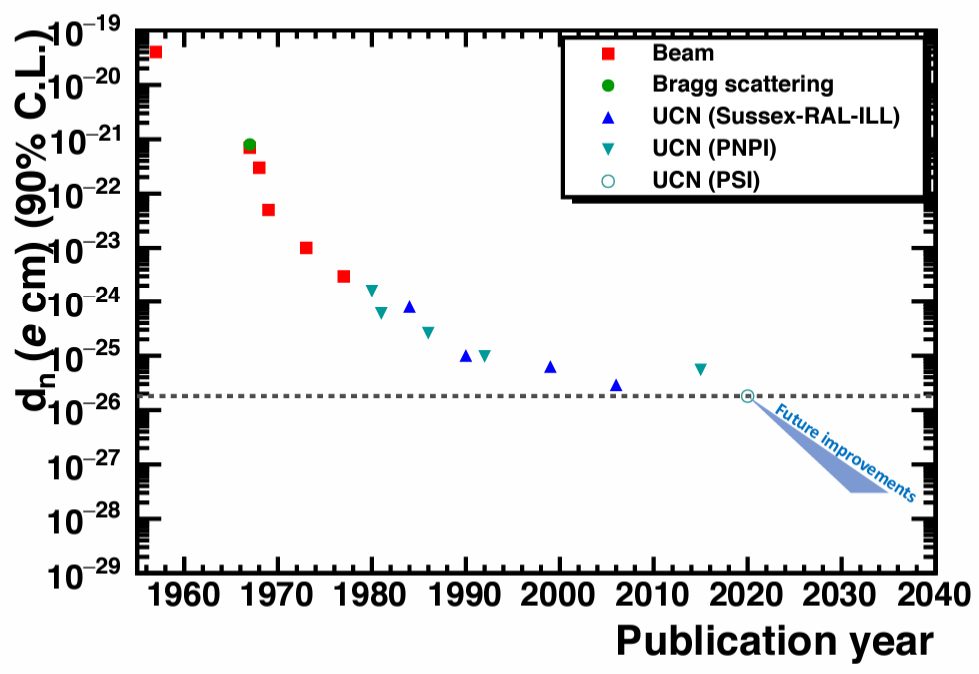}
    \caption{nEDM experimental progress~\cite{Snow22}.}
    \label{fig:snowmass}
  \end{minipage}\hfill
  \begin{minipage}{0.48\textwidth}
    \centering
    \includegraphics[width=0.95\linewidth]{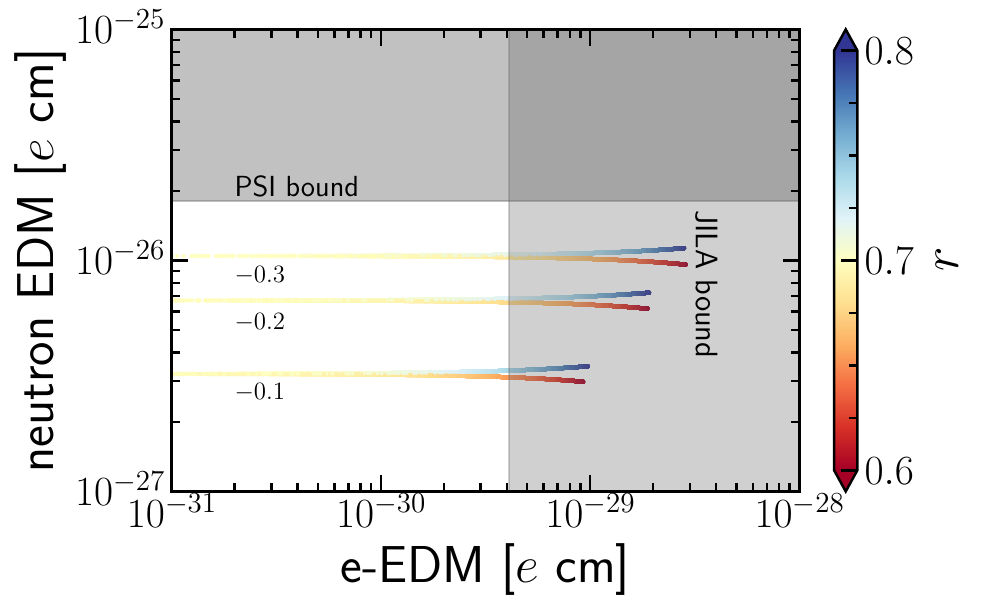}
    \caption{Combined eEDM-nEDM result.}
    \label{fig:nEDM-fixed}
  \end{minipage}
\end{figure}

Following eEDM, we turn our gaze from leptons to quarks, which brings us to neutron EDM (nEDM).
The current bound for nEDM is not as precise as eEDM, with results from PSI~\cite{PSI-nEDM20} in 2020 setting the bound at \(|d_{n}| < 1.8 \times 10^{-26} \) \(e\,\mathrm{cm} \).
Progress on the nEDM front has stagnated for a decade or so, as can be seen from Fig.~\ref{fig:snowmass} taken from the recent Snowmass report~\cite{Snow22}.
However, projects to improve the sensitivity are already in the works, so it is still worth to explore the nEDM parameter space.
\newpage

For nEDM, since quarks are the particles of interest, QCD effects come into play.
This comes in the form of two additional contributions: the chromo-EDM \(\tilde{d}_{f} \) for fermion \(f \), and the Weinberg term \(C_{W} \) for gluon interactions~\cite{Weinberg89}, which are found in the Lagrangian from the operators
\begin{equation}
  -\frac{i g_{s}}{2}\tilde{d_{f}}\left(\bar{f}\sigma^{\mu\nu}T^{a}\gamma_{5}f\right)G^{a}_{\mu\nu}\quad \qquad \quad -\frac{1}{3}C_Wf^{abc}G^{a}_{\mu\sigma}G^{b,\sigma}_{\nu}\tilde{G}^{c,\mu\nu}
\end{equation}
We use the recent formula~\cite{Hisano15}
\begin{equation}
  d_n = - 0.20\,d_u + 0.78\,d_d + e\,(0.29\,\tilde d_u + 0.59 \tilde d_d) + e\,23\;{\rm MeV}\,C_W
\end{equation}
to estimate the nEDM.
We evaluate the contributions to \(\tilde{d}_{u, d} \) and \(C_{W} \) in {\gthdm} by following Refs.~\cite{Abe14} and \cite{JungPich14}, with discussion on theoretical uncertainties found in Ref.~\cite{KanetaEtAl23}.
We present combined results for eEDM and nEDM in the range \(r \in [0.6, 0.8] \) in Fig.~\ref{fig:nEDM-fixed}.
We see that, even for \(|\rho_{tt}| = 0.3\sqrt{2} \approx 0.42\), one can still survive the current PSI bound, with the eEDM cancellation mechanism at \(r \approx 0.7 \) clearly illustrated.
The follow-up project at PSI, named n2EDM~\cite{n2EDM21}, plans to reach a sensitivity of \(\sim 10^{-27} \) \(e\,\mathrm{cm} \) within a decade, which covers the range illustrated in Fig.~\ref{fig:nEDM-fixed}.

\begin{figure}[h]
  \centering
  \includegraphics[width=4.9cm,height=3.7cm]{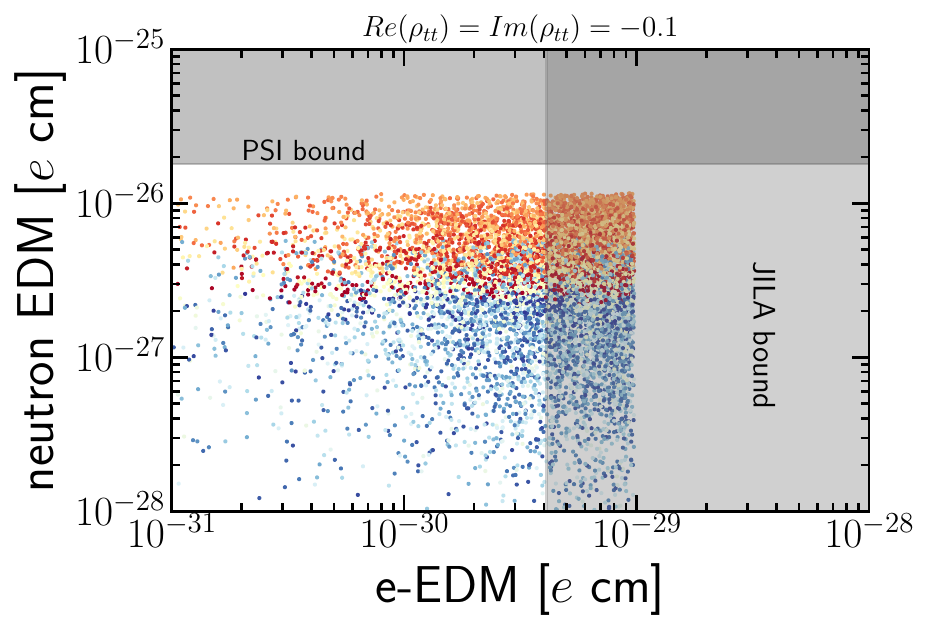}
  \includegraphics[width=4.3cm,height=3.7cm]{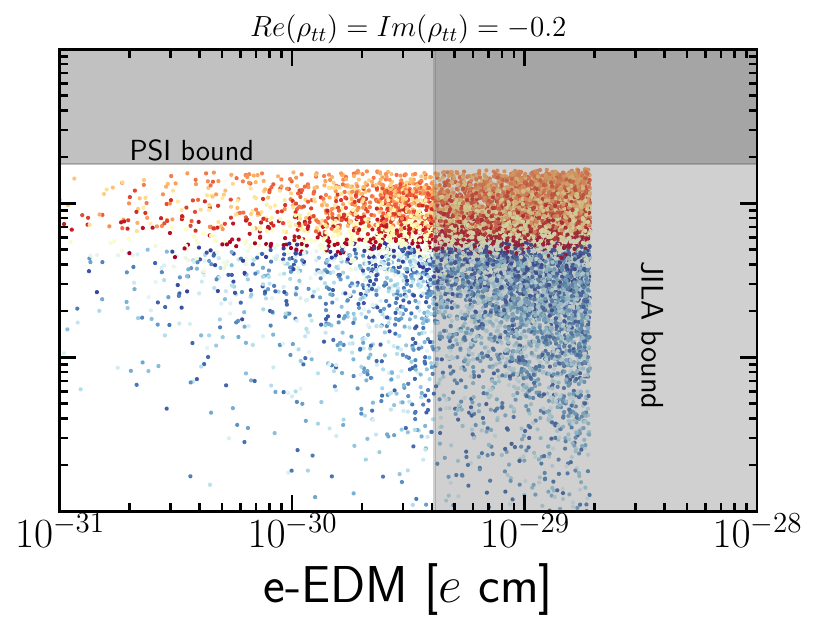}
  \includegraphics[width=5.26cm,height=3.7cm]{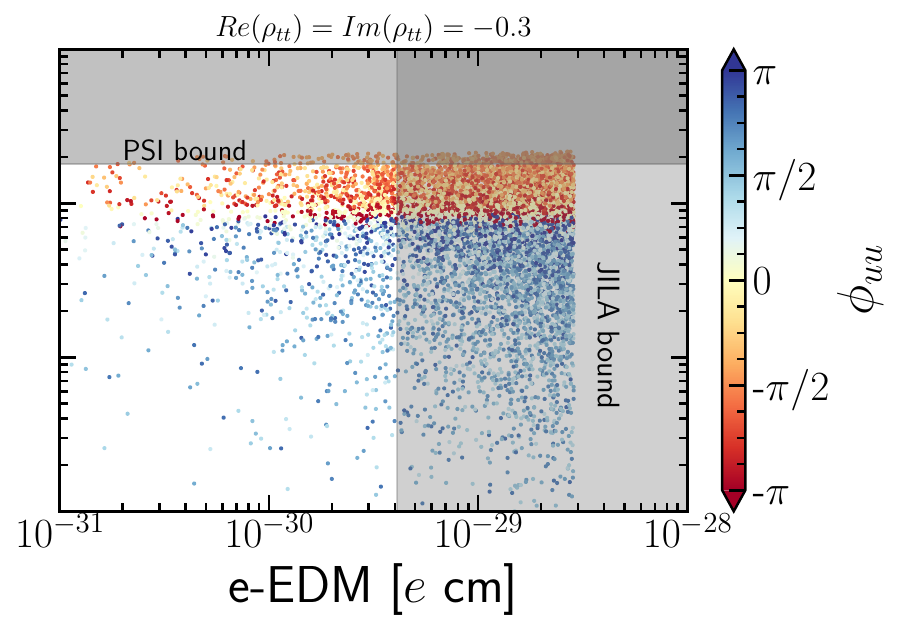}
  \caption{Results for eEDM and nEDM with \(|\rho_{uu}| \sim \lambda_{u}\).}
  \label{fig:nEDM-varied}
\end{figure}

However, we must acknowledge that the extension of the cancellation ansatz was merely a convenient means of numerical illustration of the \textit{flavor hierarchy}.
Logically speaking, since \(\rho_{uu} \) and \(\rho_{tt} \) are in the same \(\rho \) matrix, and the ansatz obviously does not hold for \(\rho_{tt} \) itself, there is no reason to expect it to hold for \(\rho_{uu} \).
Thus, we remove this ``restriction'' on \(\rho_{uu} \), and take a step back to the \textit{rule of thumb} (Eq.~\eqref{eq:ruleofthumb}). We explore \(|\rho_{uu}| \sim \lambda_{u}\) by varying
\begin{equation}
  |\rho_{uu}| \in [0.3\lambda_u, 3\lambda_u], \quad \arg\rho_{uu} \in [-\pi, \pi]
\end{equation}
while keeping the other \(\rho_{ff} \)s intact, and present some of our results in Fig.~\ref{fig:nEDM-varied}.
We see that for negative \(\arg\rho_{uu} \), i.e. same sign as \(\rho_{tt} \) (red points), the nEDM is larger but stays mostly below the PSI bound.
However, interestingly, for \textit{positive} \(\arg\rho_{uu} \) i.e. \textit{opposite} sign as \(\rho_{tt} \) (blue points), the value of nEDM drops significantly, reaching as low as \(10^{-28} \) \(e\,\mathrm{cm} \) or lower, evading even the projected sensitivity of n2EDM at PSI!
Fig.~\ref{fig:nEDM-varied} thus illustrates a \textit{natural} cancellation mechanism present within the dynamics of nEDM.
This mechanism can still be probed by future experiments, though, such as the Spallation Neutron Source (SNS) at Oak Ridge National Laboratory (ORNL)~\cite{SNS-ORNL}, which can reach sensitivities down to \(\sim 10^{-28} \) \(e\,\mathrm{cm} \).
Even though this experiment may take more than a decade to come to fruition, it almost fully covers our projected range, since the blue dots are still mostly concentrated above \(10^{-28} \) \(e\,\mathrm{cm} \).

\section{Discussion and summary}
Before stating the conclusion of our study, there are two points we would like to mention.
First, throughout this talk, we have kept the exotic Higgs masses degenerate at \(500\,\mathrm{GeV} \); however, we have explored the case of \(300\,\mathrm{GeV} \), where baryogenesis should be more efficient, and the results are similar.
The parameter space should be larger if we were to break the degeneracy of the exotic Higgs masses.
However, we would need to face electroweak precision constraints~\cite{PDG}, taking either the custodial symmetry case of \(m_{A} = m_{H^{+}} \), or the twisted-custodial~\cite{GerardHerquet07} case of \(m_{H} = m_{H^{+}} \).
Second, this study was originally motivated by the ability of the LHC to probe top CPV through top chromo-moments~\cite{CMS-tCPV}.
At the moment, the bounds on the top chromo-moments are relatively weak, so we shifted our gaze towards nEDM, which involves the up and down chromo-moments, and found out that the prospects for {\gthdm} here are rather good.

As we have illustrated in this study, with  \(\rho_{tt} \) slightly less than \(\order{1} \), eEDM below current bounds is achievable through a \textit{flavor hierarchy}-based cancellation mechanism, and the projected values for nEDM are relatively close to the current bounds, with a \(natural \) cancellation mechanism lowering the values even further.
On the experimental side, the future prospects do seem rather promising.
For nEDM, n2EDM at PSI and SNS at ORNL are expected to bring sensitivities down to \(\sim 10^{-27} \) and \(\sim 10^{-28}\) \(e\,\mathrm{cm} \), respectively, which probes most of our projected parameter space.
For eEDM, the example of the ARGUS discovery of \(B^{0}-\bar{B}^{0} \) mixing~\cite{ARGUS87} right on the CLEO bound~\cite{CLEO86} signifies that a discovery of eEDM at \(\sim 10^{-30} \), or even \(\sim 10^{-29} \) \(e\,\mathrm{cm} \) is not out of the question.
If {\gthdm} is indeed behind EWBG, the improvement of experimental precision on both fronts seem poised for discovery within the next decade or two, a \textit{double whammy} so to speak!

In summary, the \textit{general} two Higgs doublet model, without \(Z_{2} \) symmetry, can achieve baryogenesis while simultaneously evading EDM bounds through an echoing of the observed flavor hierarchy.
The improved precision of the eEDM and nEDM experiments may shake up new discussion in the realm of CPV.
Along with direct searches for exotic Higgs at LHC, ongoing efforts at Belle II as well as other flavor frontiers, we might soon see whether we can unveil what \textit{Nature} has laid out for baryogenesis.

\vspace{4ex}
\acknowledgments I thank my advisor George Hou for the opportunity to give the talk. I also thank Girish Kumar for providing much-needed guidance throughout the study. Figs.~\ref{fig:eEDM},~\ref{fig:nEDM-fixed},~\ref{fig:nEDM-varied} are taken from Ref.~\cite{HKT23-2}.

\end{document}